Vera Danilova, research fellow, PhD in Applied Linguistics, Russian Presidential Academy of National Economy and Public Administration, Moscow, Russia

maolve@gmail.com

ResearcherID AAG-2204-2021

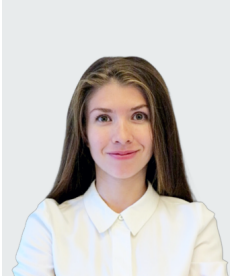

Svetlana Popova, senior lecturer at the Saint Petersburg State University, PhD candidate at TU Dublin

spbu.svp@gmail.com

ORCID: 0000-0001-5827-984X

ResearcherID: I-7008-2013

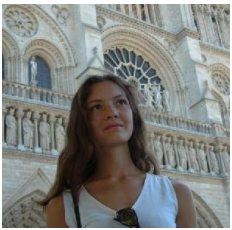

Vera Karpova, senior lecturer, PhD in Sociology, Moscow State University, Sociological faculty, Moscow, Russia

Wmkarpova@yandex.ru

ORCID: 0000-0003-2560-6140

ResearcherID AAQ-5849-2021

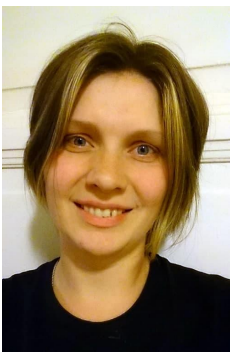

# A Pipeline for Graph-Based Monitoring of the Changes in the Information Space of Russian Social Media during the Lockdown

*Abstract*. With the COVID-19 outbreak and the subsequent lockdown, social media became a vital communication tool. The sudden outburst of online activity influenced information spread and consumption patterns. It increases the relevance of studying the dynamics of social networks and developing data processing pipelines that allow a comprehensive analysis of social media data in temporal dimension. This paper scopes the weekly dynamics of the information space represented by Russian social media (Twitter and Livejournal) during a critical period (massive COVID-19 outbreak and first governmental measures). The approach is twofold: a) build the time series of topic similarity indicators by identifying COVID-related topics in each week and measuring user contribution to the topic space, and b) cluster user activity and display user-topic relationships on graphs in a dashboard application. The paper describes the development of the pipeline, explains the choices made and provides a case study of the adaptation to virus control measures. The results confirm that social processes and behaviour in response to pandemic-triggered changes can be successfully traced in social media. Moreover, the adaptation trends revealed by psychological and sociological studies are reflected in our data and can be explored using the proposed method.

Keywords: COVID-19, topic modelling, topic dynamics, time series, user activity monitoring, LDA, feature elaboration, content analysis, social graphs

## 1. Introduction

The 2020 lockdown resulting from the outbreak of COVID-19 led to significant and sudden changes in people's lives and attitudes. In this period, social media (SM) has become a vital source of connection between people and an indispensable tool employed by governments, universities, organizations and others for information production, spread, exchange and consumption. At the same time, the informational overload went ahead feeding the online infodemic including over-interpreting, personal views, rumours, fake news, propaganda and misinformation that affected users' well-being and behaviour and threatened the effectiveness of public health measures[1] [Cinelli, 2020]. Being the main producer of the image of pandemic-related processes and events, SM undoubtedly plays the key role in their perceptions and the potential consequences thereof [Tsao, 2021] [Al-Dmour, 2020].

Researchers in different fields recurred to SM data to help the prevention and treatment procedures, as well as to explore, understand and predict the changes caused by the onset of COVID-19 and associated events [Chakraborty, 2020] [Tsao, 2020]. Text analysis tools are applied in this context, among others, to study and compare user activity across platforms, model the information and infodemic spread, find sources that are susceptible to misinformation [Cinelli, 2020], find correlations between the increase of new infection cases and public attention peaks [Hou, 2021], detection and prediction of outbreaks [Jordan, 2019].

The goal of the present research is the study of the impact of the pandemic on the online information environment in the Russian-language SM. It complements previous research on the analysis of changes in online information space around COVID-19 conducted for (mainly English-language) SM platforms including Twitter, Instagram, Youtube [Tsao, 2021] [Cinelli, 2020]. The relevance of this study derives from its key aims to understand the issues faced by people in the online environment where the information is growing and spreading uncontrollably. The investigation of information spread is particularly important during critical moments, self-isolation and lockdown in this case study.

---

[1] Managing the COVID-19 infodemic: Promoting healthy behaviours and mitigating the harm from misinformation and disinformation  (last accessed: 15.03.2021)

This article describes collection, processing and visualization of data reflecting dynamic topic-user relationships for the exploration of the changes in the Russian information space from the social science perspective. The study collects Russian-language data from two SM platforms: the Livejournal (LJ) community hosting and Twitter microblogging service. The study period starts several days before the first Address to Nation on the Coronavirus (25.03.2020) and covers the major lockdown (until June 1).

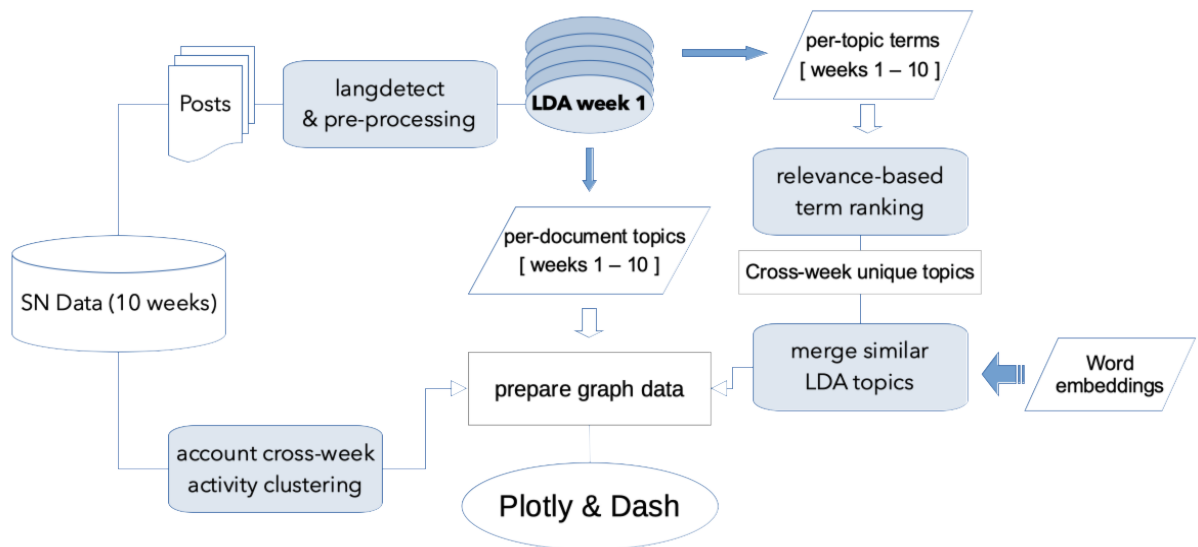

Fig. 1 Graph generation workflow

On the one hand, we study the process of discussion formation and evolution in SM during the first-ever similar unusual period: self-isolation regime was introduced in almost all regions; a non-working month was announced; social, economic and a part of political activity (the vote on the amendments to the Constitution) stopped and moved online. The sudden and drastic changes undoubtedly influenced the information spread in SM. The milestone events that took place in that period including the Addresses to the Nation by the President, the self-isolation decrees, and the restrictive measures introduced by Moscow authorities allow tracing the response of the online community. On the other hand, this study complements and expands similar research based on other online platforms providing new data to analyze information spread patterns in SM.

In the COVID-19-related studies of SM in the previous work (overviewed in the Related Work section), the use of topic modelling (mostly Latent Dirichlet Allocation (LDA)-based [Blei, 2003]) is mostly limited to getting a general view of themes in the dataset. A few works explore topic dynamics within a certain period. However, they do not consider the changes that take place in the user layer. The user dimension allows representing the size of a topic in a given time unit using not only the ratio of topic-related texts, but also the ratio of contributors. Different ways of grouping the contributors (by activity type, by dispersion) provide new insights into the study of topic dynamics and cross-network comparison of content generation behaviour. Also, to our knowledge, there is a lack of studies of public attitudes in Russian SM during lockdown.

To trace topic dynamics in connection with the user dimension and explore the changes in content generation behaviour related to COVID-19 in SM within the study period, the following pipeline is used (Fig. 1). Week is adopted as the time unit. One branch is responsible for retrieving document topics, constructing the topic hierarchy and weekly topic representations, and the other clusters account activity patterns across the weeks. The topic modelling (TM) branch solves the following main tasks: 1) for document d from the subset $D_{wn}$ published during week w in network n, identify the prevalent topic[2] t from the subset of LDA-produced topics $T_{wn}$ related to COVID-19 and active during w in n, 2) trace similar topics across all $T_{wn}$ subsets across time units in each n, 3) consolidate similar topics into larger themes and build the set T of unique topics across w and n to facilitate cross-network topic dynamics comparison. This branch produces 1) time series of topic dynamics indicators, and 2) data for graphs. Latent Dirichlet Allocation (LDA) with Gibbs sampling [Mimno,

---

[2] sets of words or phrases that tend to co-occur in texts and are associated with a certain subject, event or knowledge area

2011] is used as the base tool for topic identification. The influence of parameter tuning on the resulting topics is explored, and the optimal settings are determined based on comparative analysis and literature overview.

The other branch clusters all contributors by activity pattern. The graph data preparation step uses the output of both branches and additionally groups users by dispersion (by the number of prevalent topics in their posts). The resulting dashboard application allows producing graphs for each week, network and activity cluster by changing the corresponding settings.

The analysis based on this tool explores patterns, if any, in the dynamics of content generation behaviour across the weeks based on the obtained graphs and time series of topic dynamics. The collected data is expected to provide insights on whether the changes in the information space for both platforms are similar, what patterns are shared, or, if not, what are their distinctive features. Moreover, the gathered data is aimed to provide a better understanding of the framing of discussions around topics related to COVID-19 lockdown.

The case study (based on the Twitter data) aims to trace personal adaptation to the sudden changes and restrictive measures in SM and compare the observed pattern with the results of the available psychological studies of the adaptation during the lockdown.

## 2. Related Work

The present literature overview summarizes, on the one hand, text mining and, specifically, topic modelling techniques that were used for SM processing in the COVID-19-related research, and, on the other hand, studies that explored the impact of the pandemic on the population.

## 2.1 Text mining of SM in COVID-19-related research

The 2019-2020 events resulted in an outburst of COVID-19-related research. An extensive bibliometric analysis of officially published and indexed reports on COVID-19 collected on October 14th of 2020 [Wangtian, 2021] shows a large amount of contributions in healthcare, biology, medicine, and epidemiology. Less contributions are reported in transmission (disease transmission route), psychology and even fewer in other research directions including social impact and social science. The applications of SM data to COVID-19-related research are particularly discussed in [Tsao, 2021]. This survey summarizes 2405 peer-reviewed studies for the period from November 2019 to November 2020. It reports the use of SM mining mainly to investigate problems in psychology (mental health assessment), healthcare (detection and prediction of infection cases, evaluation of health information in prevention education videos), and social science (monitoring of public attitudes, analysis of government responses to pandemic). Also, another important research issue concerns the mechanisms of infodemics spread detection on SM. Twitter and Sina Weibo (a Chinese microblogging website) are the most frequently studied platforms [Tsao, 2021].

TM techniques (mainly LDA) are used to analyze public attitudes and concerns in English-language SM [Boon-Itt, 2020][Jelodar, 2020][Kurten, 2020][Xue, 2020][Wicke, 2020]. Some works combine LDA with sentiment analysis tools to measure public sentiment towards particular topics [Das, 2020: 163][Abd-Alrazaq, 2020]. In [Boon-Itt, 2020] and [Kurten, 2020], LDA is used to obtain a general view of the discussion topics within the studied period; topic dynamics is not covered. Manual analysis of per-topic per-word distributions is performed. [Boon-Itt, 2020] group LDA topics into larger themes by means of qualitative content analysis. These works employ sentiment analysis tools separately from TM, therefore, user sentiments towards the topics cannot be traced. [Xue, 2020] applies LDA on a large dataset of English-language tweets (April 12 to July 16, 2020) to trace public discourse on family violence. As in [Boon-Itt, 2020], similar topics (50 in total) are manually grouped into 9 larger themes followed by a detailed interpretation. [Wicke, 2020] explore both less and more granular LDA topics in the dataset by setting the number of k to 4 and 16, respectively. The topics are interpreted and summarized based on the manual analysis of lexical units. They are further used to check and analyze the presence of figurative frames, such as WAR, and the conventional metaphor DISEASE TREATMENT IS WAR in COVID-19-related themes on Twitter in March and April 2020. The above studies do not consider the temporal and user dimensions.

[Medford, 2020] considers topics and sentiments separately, however, they are further grouped with respect to the percentage of top retweeted tweets to show the prevalent sentiment and top-3 topics in each group. In [Das, 2020: 163], two LDA models are built for two parts of the dataset previously classified by sentiment using the R-based software package 'sentimentr' and the NRC Emotion Lexicon. Word cloud visualization of 20 topics from each sentiment group is made to provide insights on the contexts that are characteristic of each group. The resulting topic coverage trends for both sentiments turned out to be very similar. [Abd-Alrazaq, 2020] first performs TM of tweets published between February 2 and March 15 followed by sentiment analysis (Python textblob library) and interaction rate calculation for each of the topics based on the number of retweets, likes and followers. The number of topics that represent people's concerns in Twitter is selected manually based on the LDA output and n-gram clouds examination.

Some of the overviewed studies conducted time series analysis. [Das, 2020: 158] and [Boon-Itt, 2020] explore the dynamics of sentiment indicators. In [Das, 2020: 158], daily dynamics of sentiment in tweets tagged with \#IndiaLockdown and \#IndiafightsCorona are analyzed for the period from March 22 to April 21, 2020. Here, the daily sentiment is measured using both the whole corpus and individual tweets. [Boon-Itt, 2020] analyze the changes in the number of retweets and likes in a corpus of 107.990 English tweets related to COVID-19 between Dec 13, 2019, and March 9, 2020. The peaks are aligned with the events that were reported in the news media in the corresponding period. [Kurten, 2020] investigate daily changes in the number of English, Dutch, and French tweets and retweets posted in Belgium within the period from Feb 25 to March 30, 2020 and associate the peaks with the events that took place in this country in the corresponding period.

The work by [Gozzi, 2020] is closer to ours in that it explores topic dynamics. The base 64 topics are extracted from a corpus of news outlets published from February 7 to May 15, 2020. Temporal changes in the attention of Reddit users to the 64 news topics are measured by tracing the corresponding topics in Reddit comments using LDA for the period from February 15 to May 15. The authors notice that the view of Reddit topics is limited, since only news-triggered discussions are covered. [Medford, 2020] visualize LDA-based topic dynamics from January 14 to January 27 using a t-distributed Stochastic Neighbor Embedding (t-SNE) graph.

Following the mentioned researchers, we use LDA for topic modelling in our task. Expanding the overviewed works that predominantly process English texts and focus on the dynamics of keywords, retweets and sentiment [Das, 2020][Boon-Itt, 2020], our research, being closer to [Medford, 2020][Gozzi, 2020], collects features to explore the temporal changes in two dimensions of topics and users in Russian SM. It allows exploring the content generation behaviour within the critical period of strict lockdown and self-isolation and performing cross-network comparison.

## 2.2 Research on the Impact of the COVID-19 Pandemic on the Population

Nowadays, there is a growing body of research that studies social and psychological effects of the COVID-19 pandemic on the population and people's adaptability to a new way of life [Xiong, 2020][Chu, 2020]. Many psychological studies conducted during the pandemic, and primarily during the lockdown, explore the changes in the psychological state during the confinement starting from the establishment of lockdown measures and in 1-2 months. The results show that within the first 1-2 weeks anxiety levels rise sharply and then over the next 4-6 weeks they gradually decrease to return to their original level [Dali, 2021: 606].

In general, psychological and social consequences of the COVID-19 pandemic and restrictive measures introduced around the world are subject to further research. However, the first large-scale literature surveys show that the consequences for well-being and mental health may be quite severe. So, the authors of the overview in [Clemente-Suárez, 2020: 9] mention such manifestations as anxiety, panic attacks, depression, signs of PTSD and even suicidality. These negative consequences are aggravated by social isolation, a forced decrease in physical activity and social contacts, as well as grief after the death of friends, spouses or relatives.

As noted in [Ruggieri, 2021], one of the effective ways of coping with the experience of restrictions caused by the lockdown is online communication. The results of some studies confirmed that online communication and comparing one's own experiences of social isolation with other people reduces negative psychological consequences of quarantine [Ruggieri, 2021]. Thus, the analysis of SM posts can be a valid way to examine the coping strategies with respect to the effects of the COVID-19 pandemic and lockdown measures, since this analysis is based on the product of these experiences. This type of interaction in the absence of live communication generated collective experience of social upheavals associated with the imposed restrictions and alarming news.

There is a large-scale study currently being conducted in the US [Rusch, 2021] on a similar topic (coping during COVID-19). One of its methodological tasks is to combine the greatest number of qualitative methods and approaches to deal with the great diversity of changes observed in the public sphere during the pandemic and lockdown. However, many of the difficulties faced by this kind of research during textual data processing can be solved automatically. For instance, one of the important parameters is the frequency of polls. In text processing, the analysis can be carried out without interruption and the duration of analysis stages can be set to any value from 1 day to one week or even a month. Longitudinal surveys are particularly challenging because they are conducted multiple times. In SM mining, the total duration of the study will be primarily limited by computational performance and researcher's decisions.

Another methodological advantage of this study is the generation of a large number of questionnaires for different topics, which enables tracing specific short-term (local) topics that are discussed during 1-3 weeks and then dropped. Local topics represent issues that are relevant for the audience during a short time period. In this setting it is impossible to launch a survey with a tailored questionnaire or interview, however, we can perform text mining and identify/trace the corresponding topics.

The alignment of SM content with the main news feed for the corresponding period shows, on the one hand, that there are news stories that were definitely mirrored in SM (according to our examination of the first three weeks, the peak of discussions occurs 2-3 days after the event). On the other hand, if some news is not reflected in the SM, it means that it did not cause a significant response from the population.

Another important issue is the spread of false news (fake-news) and different manifestations of conspiracy theory [Tagliabue, 2020] that negatively affect the audience by increasing the levels of anxiety and stress. The identification of news content on SM contributes to the detection of those fake-news that get the greatest response from the audience and must be blocked in the first place.

### 3. Methodology of Feature Generation

This section discusses the approaches to the generation of topic- and user-specific features. It starts with a brief description of LDA as the base feature extraction technique explaining the choice of parameter settings and tools.

## 3.1 Latent Dirichlet Allocation

This study employs a conventional topic modelling scheme based on Latent Dirichlet Allocation (LDA) [Blei, 2003]. As indicated by the authors in [Hagen, 2018], properly trained and evaluated LDA-based topic models are a powerful tool for content analysis in social science that help discover themes overlooked by human coders and are less bias-prone. LDA is a particularly popular parametric approach that models documents as mixtures of topics and topics as mixtures of words (probabilistic distributions over words).

An overview of recent papers reporting the application of two widely used LDA-based packages, namely Java-based Mallet[3] [Zhou, 2021][Fang, 2021][Cho, 2020] and Python-based Gensim[4] [Porter, 2018][Kastrati, 2020][Riesener, 2019], as well as extensive experiments with both packages on the 1st week data (LJ) followed by the analysis of LDAvis [Sievert, 2014] output, we settled on the use of the Mallet package [Mimno, 2011]. [Ebeid, 2016] compare both tools and point out that both have their strengths and weaknesses. Mallet's underlying approach relies on Gibbs sampling, which has well-known implications for the runtime complexity [Jelodar, 2020: 2736], because the training process requires keeping the entire dataset in memory. On the other hand, as shown by [Zhou, 2021], Mallet performs better than Gensim from the perspective of the coherence value. Roughly, coherence reflects the degree of mutual support between subsets (word sets) within each topic in a topic model. C_v coherence used in this paper combines the indirect cosine measure with the NPMI (Normalized Pointwise Mutual Information) and the boolean sliding window and it is reported to be the best measure in terms of runtime and correlation to human ratings [Röder, 2015].

---
[3] http://mallet.cs.umass.edu/ (last accessed: 14.03.2021)
[4] https://radimrehurek.com/gensim/ (last accessed: 5.05.2021)

Dataset pre-processing for LDA includes lemmatization with PyMystem3[5], Russian stopwords removal using NLTK[6] and bag-of-words representation using Gensim libraries.

The best LDA setup is found as follows. Following [Wallach, 2009] we use asymmetric alpha (prior for topic proportions within documents), which combined with symmetric beta (prior for word weights in topic distributions) proved to enhance the quality of topic models. In Mallet, alpha can be optimized each N iterations using the optimize_interval parameter equal to N. In the field, it was observed that although frequent optimization increases coherence, it influences topic quality due to the growing prevalence of topics with small coverage (topics that are present in few documents)[7]. Since we aim to capture the most prominent topics, less frequent optimization is given a priority. The optimal number of topics for each week is identified by maximizing the value of c_v coherence over the following parameters: topic number in the interval from 2 to 50 and the optimize_interval in (10, 50, 100, 500, 1000). The plots of the corresponding coherence values are examined. A wider interval is avoided, because, as observed in [Porter, 2018], selecting too many topics leads to overfitting. Also, a low number of topics ensures the explicability and efficient analysis of each topic. We consider the best number of topics as corresponding to the best c_v coherence value as it is done in a number of works including [Zhou, 2021] and [Fang, 2021]. Also, in case a coherence peak that includes the highest coherence value is shared by the LDA runs with different optimize_interval values, it is considered to indicate the best number of topics. [Ebeid, 2016] and [Fang, 2021] exploit the same idea by comparing the results with varying seed. Since coherence tends to grow with the increase of the number of topics [Hasan, 2021: 350], the maximums (particularly, thosen shared by most optimize_interval values) at the beginning of the interval are given a priority. Additionally, we calculate coherence values in the same way for an extended reference corpus to compare the maximums. In this study, we identified the optimum number of topics for the first week using this method as equal to 13. Fig. 2 shows coherence peaks for the first week for the number of topics in range 5-25 (with and without an external corpus for comparison). The same procedure is conducted for the subsequent weeks. The resulting maximums are 13 or close to 13 in most weeks (9, 10, 12, 14, 16). We decided to adopt 13 as the optimum topic number for all weeks to explore the topics that come to prominence in each week and facilitate the examination of changes in the topic space. Also, we built two LDA models for the whole period to compare the unique topics, which is discussed further in the Topic Hierarchy subsection.

---

[5] https://pypi.org/project/pymystem3/ (last accessed: 20.03.21)
[6] https://www.nltk.org/ (last accessed: 5.05.2021)
[7] https://dragonfly.hypotheses.org/1051 (last accessed: 20.04.2021)

Following [Porter, 2018], we use the relevance metric with $\lambda = 0.6$ as proposed in the original paper [Sievert, 2014] to enhance the quality of topic interpretation and grouping. Relevance allows taking account both topic-specific term frequency and exclusivity under a given topic. $\lambda$ denotes term's probability weight relative to the term's lift, the ratio of a topic-specific term's probability to its marginal probability across the corpus [Sievert, 2014].

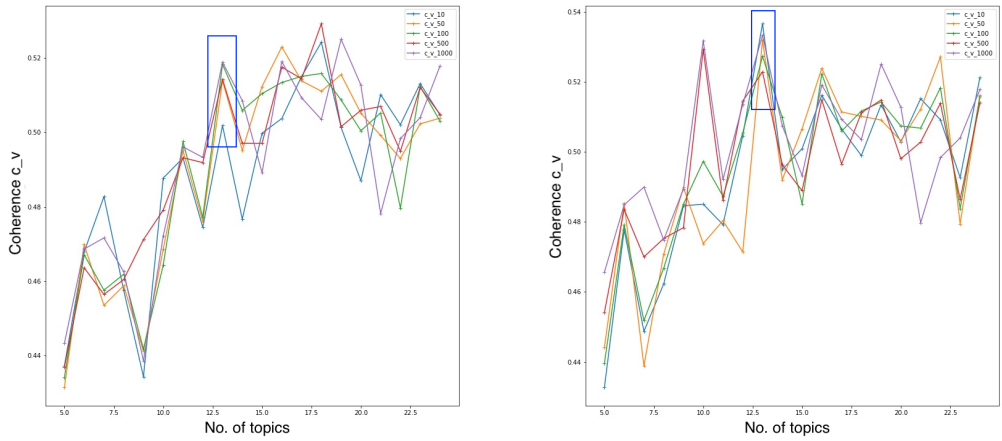

Fig. 2 Week 1 coherences for optimize_interval in (10, 50, 100, 500, 1000) with an external corpus (left) and without it (right)

In the standard output, the corpus common terms dominate representation rankings of multiple topics in the model. In this work, the main function of relevance-based ranking is to enhance the comparison of topic representations across the weeks.

### 3.2 Topic Dynamics Features

Topic dynamics features are defined in the present study as the quantitative and qualitative representations of each COVID-19 related theme that are specific to each of the time units within the considered period. The generation of topic dynamics features for both networks is achieved by: (i) pre-processing of textual data, (ii) LDA parameters selection, (iii) construction of topic models for a predefined number of topics for each week, (iv) creation of a set $S_u$ of unique topics, (v) consolidation of similar topics into larger themes to facilitate cross-week analysis and comparison of SM, (vi) iterative similarity calculation between each topic in $S_u$ and each topic in a given week's topic set to identify the changes in the representations of $S_u$ topics across the weeks, (vii) calculation of document and user-related statistics. In the following, we describe data sources, pre-processing and feature construction.

3.2.1 Data Sources

Publicly available data from the Russian version of Livejournal (LJ) and Russian-language Twitter tagged with the word "coronavirus" for the period from March 22 to June 1 of 2020 that covers the strict lockdown in Russia.

LJ is the largest online community in the Russian-language Internet that hosts the majority of the Russian top blogs[8]. According to the news media reports[9], in 2019 its audience was around 12 million people. Despite not being the largest in terms of the number of authors and messages, it is considered to be a strongly connected blogging community with a rather constant audience. As reported elsewhere, the audience is mostly male, politically oriented and 35+.

The Russian Twitter is reported to have the most active audience as compared to other SM used in Russia with an average of 47,1 messages per each of 690 thousand active authors[10]. The audience is also predominantly male (66%) and 35+ (60%)[11].

3.2.2 Textual Data Pre-processing

The basic dataset pre-processing encompasses the following steps: 1) for each sentence, the language is detected using langdetect[12], non-Russian sentences are removed; 2) empty/image posts are excluded (for LJ, empty full texts are substituted with titles, if any); 3) URLs are removed; 4) for each post, lemmatization is performed with pymystem3; 5) punctuation and parts of speech except for semantically loaded parts of speech, such as verb, noun, adjective, adverb, adverbial pronoun, interjection, numeral adjective and compounds are excluded (numeral, particle, conjunction, preposition, substantive pronoun). This is done to enhance the topic modelling performance by preserving the most semantically loaded words; 6) standard Russian stopwords are removed using NLTK. Next, the dataset is divided into 10 weeks: week 1 ("2020-03-22 - 2020-03-29"), week 2 ("2020-03-29 - 2020-04-05"), week 3 ("2020-04-05 - 2020-04-12"), week 4 ("2020-04-12 - 2020-04-19"), week 5 ("2020-04-19 - 2020-04-26"), week 6 ("2020-04-26 - 2020-05-03"), week 7 ("2020-05-03 - 2020-05-10"), week 8 ("2020-05-10 - 2020-05-17"), week 9 ("2020-05-17 - 2020-05-24"), week 10 ("2020-05-24 - 2020-06-01").

---

[8] https://www.livejournal.com/about/ (last accessed: 10.03.2021)
[9] https://www.kommersant.ru/doc/3855808 (last accessed: 10.03.2021)
[10] https://br-analytics.ru/blog/social-media-russia-2020/ (last accessed: 10.03.2021)
[11] https://lpgenerator.ru/blog/2020/01/21/10-faktov-iz-statistiki-twitter-o-kotoryh-stoit-znat-v-2020-godu/ (last accessed: 10.03.2021)
[12] https://pypi.org/project/langdetect/ (last accessed: 20.04.2021)

3.2.3 Relative word frequency and text length

To enhance topic modelling results we performed additional pre-processing of the datasets (LJ and Twitter). For each network, we identified the outliers in terms of relative word frequency that are persistent across all weeks. We explored the influence of these outliers, as well as too short and too long texts (1st and 5th quantile of length in each week) on the LDA output for the first week (in LJ). The words with significantly higher frequencies (in terms of the distance from the upper border of the main group of closely spaced frequencies) are expected to dominate the top of per-topic word distributions and multiple topics and will skew the inter-topic distances in the LDAvis visualization. The same LDA settings are used for the experiments.

***Relative Term Frequency***. The relative frequency of terms is calculated to find possible stopwords (outliers). Per-week word frequency estimation takes into account the per-document term count and the number of documents in each week. The equation is as follows:

$$f_w = 1/T * \sum_1^d n_d/N_d$$,

where n is the number of term occurrences in document d, N is the length of d and T is the size of the corresponding week's dataset (Fig. 3). The analysis of the frequency distribution shows that, for both LJ and Twitter, the term "coronavirus" has much higher frequency across all weeks. In LJ, the word "person" is another outlier. The removal of outliers increased topic coherence (from 0.46 to 0.47) and improved topic distances in the LDAvis output, which was to be expected, since, in this case, they do not help distinguishing between the topics.

*Document length*. When we consider LJ posts, the influence of document length on the topic modelling output is particularly important, because the lengths vary on average from 1.1 to 2914.5 words across the weeks in our dataset. The model assumes that documents are mainly mono-thematic, therefore it attempts to assign as few topics as possible to each text. Too long texts are likely to dominate word occurrence statistics and create too general topics.

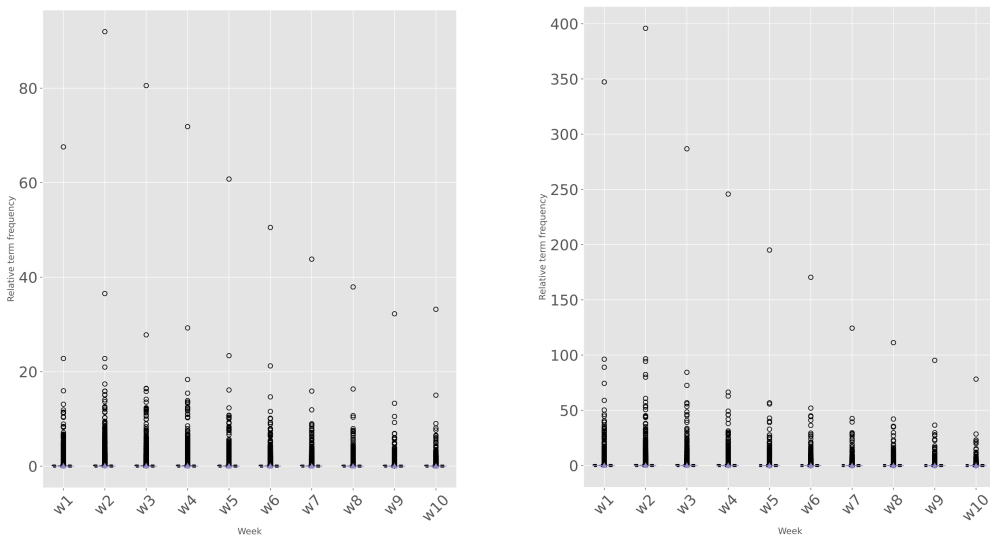

Fig. 3 Relative term frequency across weeks in LJ (left) and Twitter (right)

Too short texts tend to be ambiguous and belong to a variety of contexts, which may skew per-topic document distributions and deteriorate topic modelling quality. The length influence is examined by calculating weekly and mean values of the 1st and 5th length quantiles and comparing coherence values and LDAvis outputs for the first week's full dataset and its pruned version (without 1st and 5th quantiles).

The exclusion of the longest and shortest texts provides an insignificant increase in coherence from 0.4704 to 0.4705. However, the fact that it changed the inter-topic distances and topic contents (a new topic appeared, per-topic word distributions changed in the visualization) led to the decision to exclude too short/long texts from the dataset for further experiments. In the Twitter case, one-word texts are excluded.

The total number of posts in the final pre-processed version of the LJ dataset is 23925, the pre-processed Twitter corpus includes 66616 tweets. The dynamics of the number of posts in both networks across the weeks is shown in Fig. 4 (left).

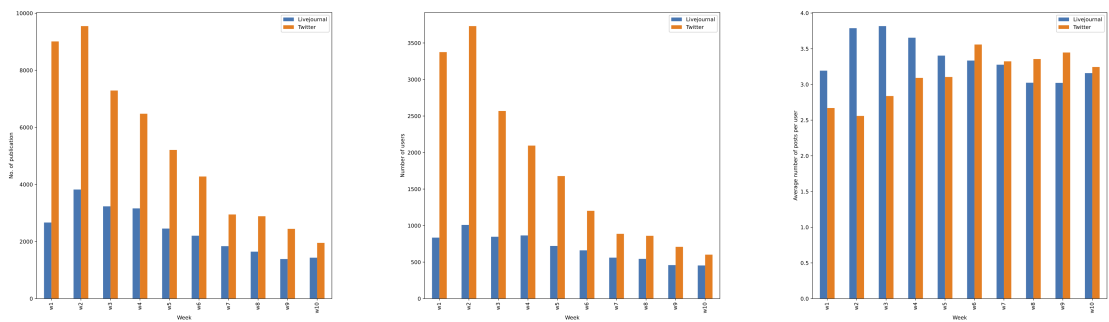

Fig. 4 The dynamics of the number of posts (left), number of users (middle), mean number of posts per user (right) across weeks in LJ (blue bars) and Twitter (orange bars)

3.2.4 Construction of the unique topics set

For each week and platform, per-topic term distributions are ranked by relevance and the list of all different topics (base list) is built as follows. First, week 1 LJ topics are assigned to the base list. Next, a pairwise comparison with the topics from weeks 2-10 is performed. If the intersection between the lists of top-50 relevant topic terms in a pair is equal or exceeds 30%, the topics are considered similar. The topic whose intersection with the base list topic is below 30% is appended to the base list. In this way, an iterative comparison and base list extension is accomplished. The resulting base lists include 42 topics (LJ) and 94 topics (Twitter).

3.2.5 Topic hierarchy construction

The construction of sets of unique topics for LJ and Twitter resulted in an overall number of 136 unique topics for both networks. On the one hand, it is to be expected that Twitter discussions are more dynamic and the topics are more diverse due to a large number of authors with their characteristic vocabularies. In the LJ community, groups of contributors develop and maintain a more 'constant' set of topics along the weeks. On the other hand, manual examination of topic content showed that within Twitter there are quite a number of topics that can be grouped into larger themes. Groups of short tweets cover only certain aspects of the same theme and LDA 'sees' them as individual topics. Also, Twitter per-topic term distributions ranked by relevance differ from LJ ones in that, in Twitter, only top-15 terms clearly define the topic, while in LJ all 50 terms are semantically connected. Moreover, we observed that most unique topics are similar across networks and represent parts of broader concepts.

To obtain a topic hierarchy and facilitate cross-network comparison, we tried both automatic (semantic clustering based on a specially trained skipgram model[13]) and manual topic consolidation. In this work, we settled on the use of the manual gold standard version, since the semantic clustering approach needs additional improvements. The manual

---

[13] https://radimrehurek.com/gensim/models/word2vec.html (last accessed: 17.07.2021)

clustering is made by two experts and assisted by a sklearn t-SNE[14] visualization of BERT[15] embeddings of the unique set topics' representations.

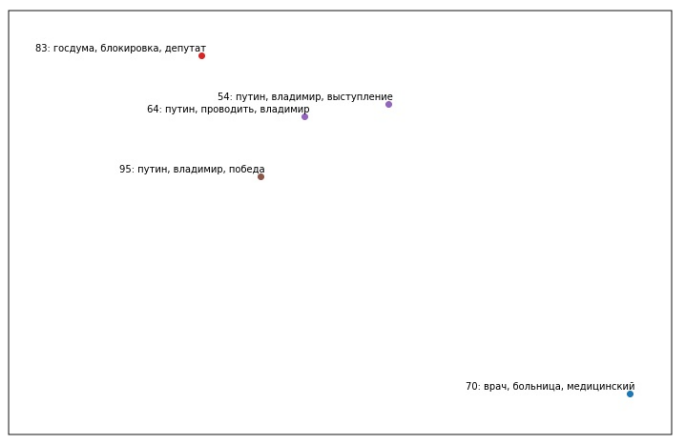

Fig. 5 An example t-SNE plot of the distribution of LDA topics from the unique topics set

The t-SNE algorithm uses dimension reduction to produce 2D scatter plots (Fig. 5) of the distribution of high-dimensional objects, such as word or document embeddings. Most dissimilar objects are separated by larger distances.

E.g., in Fig. 5, the topics related to governmental authorities (83: State Duma, blocking, deputy; 54: Putin, Vladimir, speech; 64: Putin, hold a speech, Vladimir; 95: Putin, Vladimir, Victory) are in the left upper corner and in the lower right corner there is an LDA topic that belongs to the theme "Hospitalization" (70: physician, hospital, medical). The resulting 18 large themes together with the logic of their formation (high-relevance terms that define each theme) are given in Appendix A.

3.2.6 Time series of topic-specific indicators

Similar topics are traced by performing an iterative comparison of the unique topics set with the LDA topics from each week in the corresponding datasets. In case there are > 1 topics similar to the unique set topic, all of them are included in the time series of cross-week

---
[14] https://scikit-learn.org/stable/modules/generated/sklearn.manifold.TSNE.html (last accessed: 20.07.2021)
[15] http://docs.deeppavlov.ai/en/master/features/pretrained_vectors.html (last accessed: 20.07.2021)

topic similarities. The resulting datasets for LJ and Twitter contain, for each topic t, a time series $s_t$ of topic similarity indicators representing the evolution across the study period. For a given topic, the indicators include median intersection relevance (if several similar topics are found, it helps to identify the closest one), the ratio of topic-specific texts to all week texts, the ratio of topic contributors to the total number of unique users in a given week, ratio of one-topic contributors to the total number of unique users in a given topic, ratio of one-topic contributors to the total number of one-topic users, as well as the shifts in term composition.

### 3.3. Generation of User-specific Features for Graph Visualization

The account data is grouped in two ways. On the one hand, the cross-week account activity data (number of texts per week) is clustered using k-means to explore activity patterns. Particularly, we test the assumption that there are users sharing similar activity patterns (in per-week per-user number of posts) within the study period. The following groups are expected: 1) users whose period of maximum activity lasts one or two weeks, which may reflect their response to some impacting news or events or purposeful information spread in case of a large number of messages; 2) users who contribute uniformly or almost uniformly during the study period (the cluster of the most active users, these are often news accounts); 3) users whose interest in COVID-19-related topics gradually fades from the 1st to the 10th week.

On the other hand, we group users by dispersion (participation in n topics) to explore the per-topic contribution of the corresponding groups during a given week.

#### 3.3.1 Account Activity Clustering

The total number of users that contributed during the study period is 14129 (Twitter) and 2470 (LJ). The general trends in weekly dynamics of the number of users and the mean number of posts per user are depicted in Fig. 3 (middle and right), respectively. In both networks, the activity decreases, in Twitter, a more drastic change is observed, while the mean number of messages per user lies between 2 and 4, which is most likely due to a large number of users with very few messages per week in both networks.

The clustering is performed using k-means from Python-based sklearn library[16]. The matrix of per-user per-week messages is normalized by rows (by the overall per-user contribution) and by columns using StandardScaler from sklearn. The number of clusters is selected using the popular elbow curve method [Marutho, 2018] based on the within-cluster sum of square distances. Based on the experiment with 9 different random seed values and the number of clusters in range (2, 15) the optimal number of clusters is determined as equal to 11 for LJ and 10 for Twitter. The percentage of users per cluster in LJ and Twitter is shown in Table 1. The clusters are visualized in Fig. 6. The colours correspond to cluster numbers that are identical to those shown in Table 1.

Both in LJ and Twitter, there are groups whose largest contributions are made in one of the weeks with none or almost no activity before and after the corresponding week. Also, the largest groups (cluster 3 in LJ and 0 in Twitter) are represented, among others, by the users who were the most active (or uniformly inactive) along the whole period.

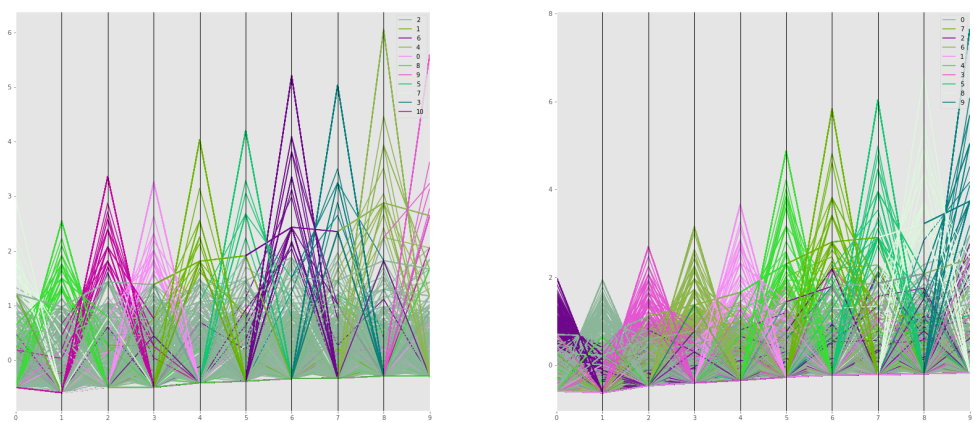

Fig. 6 User activity clusters: 11 clusters in LJ (left) and 10 clusters in Twitter (right)

---
[16] https://scikit-learn.org/stable/modules/generated/sklearn.cluster.KMeans.html (last accessed: 20.04.2021)

| № | 0 | 1 | 2 | 3 | 4 | 5 | 6 | 7 | 8 | 9 | 10 |
|---|---|---|---|---|---|---|---|---|---|---|---|
| Users %, LJ | 4.42 | 7.13 | 4.59 | 36.55 | 3.59 | 4.01 | 6.84 | 3.46 | 9.18 | 7.47 | 12.77 |
| Users %, Twitter | 31.04 | 7.13 | 21.24 | 13.06 | 4.78 | 2.98 | 11.77 | 3.47 | 2.59 | 1.94 | - |

Table 1 The percentage of users per cluster in LJ and Twitter

In Twitter, this group also includes users with the contribution maximum at week 2 followed by a slight decrease. In LJ, the general trend in this group is the same, however, there is an independent cluster 1 (Fig.6) with the maximum at week 2 preceded and followed by zero or almost zero activity.

The examination of user activity and attention shows consistent and predictable dynamics: gradual decline in user activity after the second week. This trend can be due to the fact that it was in week 2 from March 29 to April 5 when the introduction of the self-isolation regime from March 30 was announced as well as the start of a non-working month from March 30. In week 2, the response is the highest and it gradually fades as people are getting used to new circumstances and start creating the appropriate behavioral patterns.

3.3.2 Account Grouping by Dispersion

We tested three approaches to summarize user data given the topic distribution per user. Two previous graph versions displayed, for each week, full graphs of users and all different dispersion groups, respectively. By dispersion we mean the length of topic distribution per account. The main function of full user graphs was to display each user's "interest" in a given topic (ratio of his topic-specific texts to his total contribution) and the ratio of his per-topic publications to the per-week per-topic contribution of all users. When exploring this graph type, we noticed that, surprisingly, certain news accounts turned out to be the most 'interested' in small-sized specific topics within the theme "Entertainment and Leisure", according to LDA per-document topic prevalences. We then weighed user dispersions by the use of n top relevant terms in a given account's documents. With n == 5 the mentioned news accounts lost their positions in the "interest" ratings, since their posts did not contain top-5 relevant words. We keep working on this graph type to expand the functionalities of the current application version.

The graph of dispersion groups visualized all types of topic combinations per account in a given week where the accounts with the same topic combinations were grouped together into one node. We decided to further summarize the data, since the previous two graph types are large and difficult to interpret when performing a cross-week and cross-network analysis. We settled on summarizing the dispersion data, since it represents the audience focus properties. It is achieved by gathering all users belonging to a certain dispersion type (e.g., one-topic users) in one node (n-topic group).

### 3.3.3 Graph Visualization in Dash

To visualize the connections between n-topic user groups (user dispersion groups) and the corresponding topics in each week and cluster for each network (LJ and Twitter) we built a Dash[17] application based on the graphs created using networkxs[18] and Plotly[19]. The app allows switching between the following parameters: week (from the 1st to the 10th week), cluster (the main cluster and the one with the peak at a given week) and network (LJ or Twitter). For each week, the graph is a bidirected graph where the vertices aligned on the left are topics and the vertices on the right are user groups (Fig. 7). The information appears when the user hovers the pointer over the nodes/edges.

---

[17] https://dash.plotly.com/ (last accessed: 15.05.2021)
[18] https://networkx.org/ (last accessed: 15.05.2021)
[19] https://plotly.com/ (last accessed: 15.05.2021)

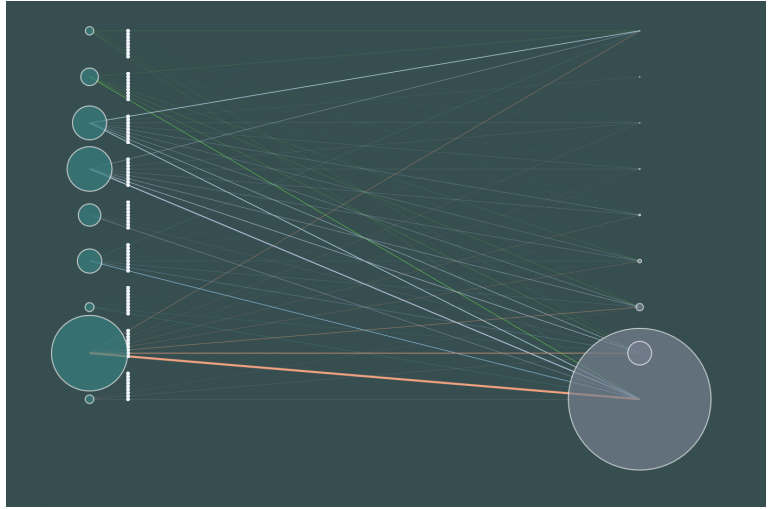

Fig. 7 A sample graph (Twitter, week 1, main cluster) of connections between user dispersion groups (right) and topics (left) for each week.

The size of topic bubbles is proportional to the ratio of accounts in a given topic in a given cluster to the total number of unique accounts in a given week. The hover information for each node shows the percentage of posts per week, the average number of posts per account, as well as the number and percentage of each n-topic group. The size of the nodes on the other side (n-topic groups) is proportional to the ratio of the number of accounts in a given group to the total number of unique accounts per week. The hover information shows, for a given group, the number of accounts, the ratio to all accounts, and the number of topics covered. The hovering information on edges (false nodes) shows the contributing group and the ratio of texts in this contribution to all week's texts. The width of edges is proportional to the ratio of contributed texts for a given n-topic users group.

The dashboard application can be run directly from the Google Colab[20]. The Google Colab page also provides links for all the obtained plots, graphs and tables.

---

[20] https://bit.ly/3tcMmuX (last accessed: 25.07.2021)

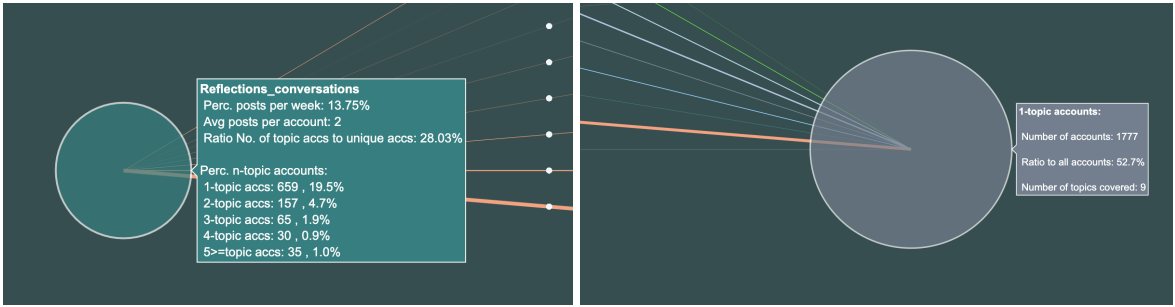

Fig. 8 Hover information on topic nodes (left) and user dispersion nodes (right)

The produced interactive graphs combine all of the data representation features for the further analysis, however, the use of their captions in the paper is not appropriate due to their size and interactive elements. Therefore, we plot the main weekly statistics for the analysis section.

## 4. Data Analysis

We perform the analysis of user activity dynamics with respect to the inferred topics for the period of 10 weeks. The correspondence of week numbers to dates is given in 3.2.2. Since the topics are directly associated with real events, we rely on a timeline[21] of these events that took place in Russia during the considered period and are found to be mirrored in SM (Table 2).

| № | Date | Event |
|---|------|-------|
| 1 | 25.03.2020 | The first Address of the President to the Nation (introduction of the self-isolation regime, non-working week is announced, 2020 Russian Constitutional Referendum is postponed) |
| 2 | 27.03.2020 | International flights were grounded after the government ordered the civil aviation authority to suspend all regular and charter flights to and from the country |
| 3 | 29.03.2020 | Mayor of Moscow Sergey Sobyanin issued a stay-at-home order starting the next day |

---

[21] https://en.wikipedia.org/wiki/COVID-19_pandemic_in_Russia#cite_note-tmt-mar29-37 (last accessed: 25.07.2021)

| 4 | 30.03.2020 | Similar orders or recommendations were announced in numerous other federal jurisdictions, with many more announcing such restrictions over the next few days. The same day, the border was shut, with all border crossings closed |
| 5 | 02.04.2020 | The second Address of the President to the Nation (establishment of penalties for the violation of the self-isolation regime, announcement of the non-working month) |
| 6 | 11.04.2020 | Moscow's mayor, Sobyanin, signed a decree introducing a digital pass system to enforce the coronavirus lockdown |
| 7 | 28.04.2020 | President's announcement of the prolongation of non-working days until May 11 |
| 8 | 09.05.2020 | Air show instead of the 2020 Moscow Victory Day Parade |
| 9 | 11.05.2020 | President Putin announced the end of the national non-working period |
| 10 | 12.05.2020 | Announcement of additional support measures |
| 11 | 27.05.2020 | Sobyanin announced that some restrictions in Moscow would be eased on 1 June |

Table 2. Timeline of the main events that took place within the considered period

Tweets are more "spontaneous": they are short (1-2 sentences) and not elaborate and polished as LJ posts. They appear more similar to spontaneous speech, reflecting more clearly the inner state of a person, his/her reactions to events and triggers, anxiety, loss of interest, joy. In the first place, this study uses the obtained data (based on the Twitter corpus) visualized on graphs to monitor the process of adaptation to new life conditions, and the corresponding dynamics of the anxiety levels. The anxiety levels are associated with the intensity of user contribution to the information space. For the purposes of this case study, we chose to focus on the Twitter data due to its nature and a large number of users as compared to LJ. In LJ (after the consolidation of LDA topics for both networks), the main activity is observed in the same topics, however, their distributions differ. The dynamics of cross-week user activity in topics in LJ and a detailed comparison of both networks are subject to future research.

The intensity of user activity is measured based on the following data: 1) the number of active users and contributions (posts) per week, 2) cross-week dynamics of topic diversity, changes in user activity and shifts in per-topic distributions of relevant words in each topic, 3) user dispersion.

Let us justify the above statements:

1. Anxiety generates the need for communication, self-expressions and discussion [Folkman, 2004]. The latter enables the increase in the number of posts/messages.

2. Two criteria are taken into account: (i) the demand for a certain topic diversity and the number of covered topics: while a user is engaged in a topic, he generates posts and actively participates in the discussions related to this topic (or its subtopics). With the decline in interest towards COVID-19, the variety of topics and the corresponding activity decreases; the topics that have the greatest impact on life are the most talked about. (ii) dynamics in the topic space contents, particularly, the transition from the topics related to one's own well-being ("self comes first") to broader contexts ("external" topics).

3. The narrowing of the thematic diversity of user-generated texts, as well as the transition from topics related directly to personal experiences to less emotionally colored and more rational topics.

Based on the analysis of the obtained data we suggest the following. The examined period is divided into 2: weeks 1-6 (the end of March and April) and weeks 7-10 (May).

**Week 1**. The growing concern amid voluntary measures (voluntary self-isolation), transition to distance learning and work, COVID-19 spread, and the first Address of the President.

**Week 2**. The first policy-induced restrictive measures that significantly affected the everyday life and household economy. It hit people like a bolt from the blue and caused a strong response from the SM audience - an additional burst of activity.

**Weeks 3-6**. Adaptation and acceptance of the introduced measures, private household issues followed by a switch to a new mode of life characterized by the transition to a broader context (online shopping, world events, aid to health professionals, etc.). A gradual drop of COVID-19-related interest rates that we associate with the decrease in anxiety levels. The turn of attention towards external topics and broader context, as well as the fading of the topic related to introspection, the need of self-expression ( "Reflections_conversations") also characterize the decrease of stress.

**Week 7**. Festive days (the Victory day and Easter )and the fading of attention towards COVID-19.

**Week 8.** The termination of non-working days and recommencement of work.

**Weeks 9-10**. The COVID-19-related topic diversity finally collapses to one major discussion on statistical reports that we associate with the complete decay of interest towards COVID-19.

Let us consider in more detail the changes in topic-related user activity that took place along the considered 10 weeks using the above listed criteria.

1. The number of active users and contributions per week (Fig. 4 (left and middle))

The highest number of active users and posts is reached at weeks 1 and 2, with a peak at week 2. Over a half of the total number of week 1 users are in the cluster of temporal activity (peak at week 1), meaning that their activity either drops or stops completely after this week. Further, starting from week 2, the number of active users drops each subsequent week until the end of April (week 6). In the following weeks, the drop either slows down significantly or stops. We assume that by May 2020 the distribution of active users across the topics approximates a standard (pre-COVID) pattern after the burst of activity in early April, so the change in the number of users is no longer as noticeable as in April. The number of active users drops for all topics inferred by the model and consolidated by domain experts. Additionally, user activity gets redistributed simultaneously with the decrease in the number of active users, which will be discussed below. Starting from 2 weeks, the percentage of active users in the topic "Statistics" begins to grow (at the expense of a decrease in the percentage of active users in other topics). Starting from week 7, the number of active users in this topic is actively growing, and by weeks 9 and 10 it becomes the only major discussion (about 80% of active users publish tweets/posts on this topic).

Thus, we observe a transition from the distribution of attention and interest between various topics to a monotopic focus. At the same time, the total number of active users is significantly decreasing, that is, over time, the number of users tweeting/posting on COVID-19 (tagged with "coronavirus") gets smaller.

2. Cross-week dynamics of topic diversity.

2.1. - Changes in user activity

The plots illustrating the analysis are placed in Appendix B due to their size. Weekly sums of account ratios displayed in the corresponding plots can exceed 100% since the same account often contributes to multiple topics and the total percentage of posts per week is always equal to 100%. Therefore the per-topic number of posts (blue bars) is often lower than the per-topic percentage of users (orange bars).

In week 1 (Appendix B.1), the highest productivity (over 23-24% of posts) is observed for the topics "Reflections_conversations" (over 40% of users), "Virus_control_measures (public level)" (over 35% of users) and "Statistics"(over 23% of users). Topic descriptions are given in Appendix A. It implies strong concern about COVID-19 and the need for discussing it. On the other hand, public protective measures are actively talked about. People are concerned with the shortage of protective means, the uncontrollable prices and the agiotage. Among other topics, household issues are the most actively discussed ("Solutions_to_household_problems", over 13% users and 6% of posts) due to the ongoing changes in everyday life and voluntary self-isolation. As mentioned above, over 50 % of the week 1 users are in the cluster of temporal activity. These are mostly

1-topic dispersion users (52.7 % of week 1 accounts) who contributed to all 9 topics giving much higher priority to "Reflection_conversations" (28% of accounts and 14% of week 1 posts) and "Virus_control_measures (public level)" (17% of accounts and 10% of week 1 posts). This spark in activity is supposed to be triggered by the first Address to the Nation that took place on March 25. This first-of-its-kind event served as an indicator of the importance and severity of the coronavirus threat, and the unprecedented measures to be introduced. The topic diversity of this activity can speak of the strength of emotional tension and general anxiety that began to spread in society this week.

In week 2 (Appendix B.2), the introduction of the first policy-induced restrictive measures sparks the crowds. Indeed, this is the largest topic discussed ("Virus_control_measures (public level)"): over 39% of the posts are assigned to it and over half of the accounts wrote about it. The week was shocking due to the introduction of a significant number of strict obligatory quarantine measures (the so-called self-isolation regime) with justification for their necessity. Such a sharp change caused an urgent need to discuss it, which is characterized by a steep increase in activity on the topic. The latter speaks of the high significance and impact of the introduced measures on people and their lives and perception of things.

The discussions on statistics reports remain almost at the same level of activity (around 20% of users and posts), which indirectly indicates the stability of this topic. During this period, the absolute number of accounts writing on the topic of statistics practically did not change, while the number of texts decreased insignificantly. In the rest of the topics, the activity is weak with "Reflections_conversations" being the most discussed (about 15% of users ), which points to the existing concern about social interaction on COVID-19 issues.

In week 3 (Appendix B.3), there remains a slightly increased participation of users in the discussion of public protective measures, although the topic's popularity declined, engaging only 15% of active users. We assume that this is a "residual effect" of the second week's maximum. It should be mentioned that the acceptance of measures as inevitable circumstances occurred already in week 1. The topic "Statistics" gains the most popularity this week in terms of both the number of accounts and posts/tweets, since the government's decisions are based on the growing number of infected people. Also, the number of texts grew faster than the number of accounts, which indicates greater per-user activity. This week, the discussion on improving everyday life ("Solutions_to_household_problems") returns and is prevalent among the rest of the topics (about 21 % of users published about 9 % of posts on it). As we observed, this topic appears at the beginning of the period, in week 1, in the context of voluntary measures, and in week 3 in the context of the introduced mandatory measures. The gap in week 2 is explained by the attention shift to new uncertain events about to occur related to the introduction of quarantine and restrictions. With the acceptance of quarantine as an inevitable condition, there is a return to vital issues, and, in the first place, to the establishment of everyday life in new conditions. Also, there is another important topic related to personal well-being in this week - work and education, including distance education ("Work").

In week 4 (Appendix B.4), after a decline in week 3 to of active users, the topic "Virus_control_measures (public level)" gained popularity again (approximately 35% of users and 22% of posts), although significantly less than in week 2. In the subsequent weeks (5-7) it remains approximately at the same level. The activity in the "Solutions_to_household_problems" drops this week. Thus, the discussion of general (public-level) restrictive measures shifts the focus of attention from solving specific practical problems of everyday life to a broader context. Also, the growth of activity in this topic appears to be due to the introduction of the system of digital permits in Moscow (the decree as of 10.04 put into effect as of 13.04). Except for these two most prominent topics (statistics and public measures), the topic "Reflections_conversations" comes back to the fore (19% of users and 11% of the posts). There appears a local (occasional) topic "Holidays" due to the Easter celebration.

This week does not feature the topics that were active during the weeks 2 and/or 3 - "Solutions_to_household_problems", "Economic_issues", "Work", "Entertainment_and_leisure", which may imply the normalization of people's understanding of the new conditions - distance learning and work, online shopping, the shift to local and more general topics.

In all of the following weeks, the dynamics of the "Statistics" topic remains the same (the highest activity rates). Therefore, the analysis describes other prominent features of the dynamics.

In week 5 (Appendix B.5), the highest activity remains in "Reflections_conversations" and "Virus_control_measures (public level)", the number of users in "Reflections_conversations" being noticeably higher (approximately 37% versus 27%). The topic "Economic_issues" reappears (approximately 8% of posts and 13% of users).

In week 6 (Appendix B.6), the topic "Reflections_conversations" drops sharply (about 8% of active posts and 18% of users). The topic of "Virus_control_measures (public level)" is still the leading one (approximately 27% of posts and 37% of users). This week the topic related to distance learning and work ("Work") comes up again, which may be due to the extension of the non-working days until the end of the May holidays (announced at the end of April).

In week 7 (Appendix B.7), "Virus_control_measures (public level)" remains the leading topic. The local topic "Holidays" returns (related to the Victory day on May 9), and a new topic appears - the support of medical workers ("Healthcare_professionals

(incl.payments)"), most likely, as a reaction to the news reports about the introduction of incentive payments for healthcare workers as of 6.05, and internal policy. This week the activity in the "Reflections_conversations" topic stops and shows up in the following weeks with the activity of about 5% of users.

By week 8 (Appendix B.8), the attention to public protective measures gradually fades covering about 17% of posts, and 26% of users, and a clear increase of the topic "Statistics" starts (64% of posts and 58% of users). The activity in other topics also declines. Among the remaining topics, peace and information are leading, which indicates a shift of interest further and further from personal problems to the general information and foreign policy level.

Week 9 (Appendix B.9) basically features 2 topics: the growing "Statistics" (85% of users and posts) and "Virus_control_measures (public level)" (10% of posts and 20% of users) whose activity keeps declining.

In week 10 (Appendix B.10), the trends of week 9 persist. Another topic that returns to the fore this week is the assistance to medical workers, "Healthcare_professionals (incl.payments)", with 22% of posts and 10% of accounts. This may be due to the shortage of protective suits and personal protective equipment observed in late May-early June in various regions of the Russian Federation.

The observed changes allow defining the main trends in the dynamics of user activity and topic contents in the examined period. In the first place, we notice the overall gradual fading of interest towards COVID-19 and the introduced measures that is manifested in 1) the decrease of active accounts and posts/tweets across the weeks and 2) the shift of the contents of the topic space towards less emotionally loaded and more rational. Secondly, news content gets quickly reflected in discussions (no longer than within 1 week, which may be due to the selected analysis unit) and rapidly disappears if the effects of the reported events are either short-term or weak (e.g., the discussions on the Addresses to the Nation and holidays keep being active for at most 2 weeks). Thirdly, the power of experience and the relevance of the COVID-19 theme to the authors is represented by the high topic diversity at the beginning of the period. Even in week 1 before the introduction of the first restrictive measures there are 9 almost uniformly active topics (except for "Statistics" and "Virus_control_measures (public level)" that dominate the activity across all weeks). By contrast, at the end of the period, with a similar number of topics, only the leading topics remain actively discussed while the coverage of the rest of the topics is on average below 10%.

2.2 Shifts in per-topic distributions of relevant words in each topic.

Let us consider the weekly changes within the main topics:

**COVID (investigation, tests, treatment)**. In the first 2-3 weeks of April, the discussions within this topic are mostly about private well-being (immune system boosting, COVID symptoms compared with flu and pneumonia, what different sources say about the virus). From week 4, the topic becomes more "detached", the attention shifts from studying the interaction with the virus at a household level to topics related to mass health, laboratory research, vaccines and the vaccine market. In the second period (May), the topic arises in weeks 8 and 9, which may be associated with a reaction to local events: first, in connection with mass antibody tests in Moscow and vaccine testing that began to be reported in the media. To summarize, during the first 2-3 weeks (until May) the topic is perceived in a more personal and emotional way and after this period it becomes more abstract and susceptible to news feeds.

**Economic_issues**. In the weeks 2-3, the discussion concerns global problems relevant to personal safety and interests: economic crisis, the closure of borders. The local context covers the governmental support for families and businesses in a difficult economic situation (weeks 1 and 3). The topic is active in the initial period (weeks 1-3), then it drops off the discussions yielding to the topics "Statistics" and "Virus_control_measures (public level)". We associate it, firstly, with the general tendency of a decrease in user anxiety and loss of interest in the former variety of topics, as well as a shift in the focus of attention to a more thematically focused context dedicated to the disease itself (its statistics and restrictive measures).

**Hospitalization**. In weeks 1-4, the topic content is quite predictable - hospitalization, patients, communal services, infection, artificial lung ventilation, etc. In weeks 5-6, the topic shifts the focus of attention: in week 5, Moscow appears in the topic and, in week 6, the topic covers the stories of hospital staff. Then the topic drops off and returns in week 8 in the context of the increase in hospital admissions in St. Petersburg, which caused an excessive burden on hospitals. The topic also reflects local events - hospitalizations of famous people. Thus, here we can also talk about a transition from more personal aspects that affect emotional stability, since there is a prospect of being admitted to a hospital, to some more detached view of what happens in hospitals.

**Information_sources**. In week 1, a greater emphasis is on the topic of distance education (schoolchildren, students) and Olympiad (Tokyo 2020). In week 2, the topic shifts towards news sources of relevant statistics and tracking the situation in the world. In the fourth week, a discussion on TV series joins in. In week 6, the emphasis is shifted even more strongly to information on leisure (films, shops, subscriptions), and the topic of distance learning comes up again. The rest of the period, it mostly contains reactions to news events (hospitalization of the Kremlin press secretary Peskov, reaction to Mikhalkov's program "Besogon", online outlet, news portals).

This topic illustrates well the turn of attention to COVID-19 in week 2 (sources of information and statistics become the focus of attention) when the virus spread starts to directly affect people's lives, pushing aside the problems of distance education. We carefully suggest that it was in week 2 that the pandemic began to be perceived as something real and impacting. Also, the emergence of interest in information on TV shows in week 4 indirectly indicates a decline in anxiety about the virus and a shift in attention to organizing leisure during the pandemic. In week 6, the leisure-related discussion intensifies and the topic of distance learning arises again, which may be related to the hope for changes after the May holidays.

**Reflections_conversations**. In week 1, the topic includes 2 parts: on one side, general discussions on the virus - its origin, current events, scientific advancements, vaccines, and, on the other side, how to spend time on quarantine with children and family, family leisure. In week 2, attention shifts to the economic and political aspects of the pandemic. In the third week, the topic of self-isolation with children returns and the topic of the economic and political aspects of the pandemic remains. In addition, week 3 discusses the virus spread and what is happening in the country and in the world (with an emphasis on China and Italy). Week 3 is the most diverse in terms of subtopics. In week 4, various foreign news feeds are discussed. In week 5, the topic related to distance education for children reappears. In weeks 5-6, the subtopic on staying home in quarantine remains. In weeks 8-10, the discussions touch on lockdown life and concerns: family time (children), job, tests, payments, medicines.

The subtopic on family life on lockdown and children is the most stable. Its importance can be attributed to a sharply increased burden on parents and difficulties in organizing children's education and leisure at home. This is associated with an increase in anxiety. This subtopic persists in weeks 1-6 (except week 2) and its tokens are among 30 with the highest ranking scores (relevance). Despite the fact that the main restrictions were announced in the second week, this discussion is strongly present only starting from week 3, which is probably due to a need to organize primary household tasks first and then there arise new problems related to the long stay in self-isolation with children and other family members in a limited space. After week 6, specific subtopics are not clearly expressed, which is assumed to indicate successful coping with the ongoing changes.

**Statistics**. In week 1, the discussions focus on the spread of the virus in the world, particularly, infection and death statistics recovered for different countries including Russia. From the second week, the global context disappears, the discussion on domestic statistics remains, Ukraine is also present. In week 2, there are statistics for Moscow and, in week 3, for the regions. In weeks 4-5, the general trend continues and world statistics reappears. Week 5 features statistics in Ukraine, Moscow, St.Petersburg, Moscow oblast, Belarus, and foreign cities. In week 7, the global context returns to the fore, the infographics and virus spread in April are discussed, and the subtopic on testing statistics appears. In week 8, there are 2 major subtopics: domestic and world statistics. In week 9, the upcoming second wave in the fall is talked about. Here, similarly to other topics, the scope narrows from global to domestic and more "personal" level through the transition to Russia, then Moscow and different regions (political subdivisions) with their specific locations. Then, from the 5th week, there is a reverse process of expanding the context with a gradual return to global statistics while domestic statistics persists.

**Virus_control_measures (personal level)**. In week 1, the discussions touch on individual protection measures (wearing a mask, staying at home, washing hands). In weeks 2 and 3, the emphasis changes to purchasing personal protective equipment ("purchase", "production", "delivery"). In week 4, a sub-discourse on the protection of contact people (workers, doctors, salespeople, etc.) arises. It persists in week 5 while the sub-topic of acquiring personal protective equipment is replaced by the sub-topic of their use. In week 6, the topic is a mixture of discourses related to purchasing and obligatory wearing of personal protective equipment in public places. In weeks 8-10, the focus is on the mandatory use of personal protective equipment in public places.

As in the previous topics, there is a transition from personal everyday practical tasks (find, buy) in the initial period to the external (protection of other people and social obligations).

**Solutions_to_household_problems**. The topic of improving everyday life is present in the first weeks of the period - up to 5 weeks. In general, the development of the topic comes from everyday issues of buying vital supplies (shops / pharmacies, buckwheat, toilet paper in week 1) through the organization of home leisure with children (3 weeks) and then goes to online leisure and shopping (5 weeks). The latter may indicate adaptation to a new lifestyle by the 5th-6th weeks and even readiness for long-term quarantine, when season tickets become relevant.

**Virus_control_measures (public level)**. In the first week, the sub-discourses are related to the President's Address, the introduction of a non-working week with salary retention, measures in Moscow, border closure and restrictions on movement. In the second week, the topic as a whole reaches its peak. It discusses the second Address of the President, the extension of non-working days, the introduction of the self-isolation regime, the closure of most stores, restrictions on visits to grocery stores, pharmacies, self-isolation fines, support measures, and spending time at home on lockdown. The self-isolation discourse (staying home, family relations) remains stable in weeks 3-7. In week 3, discussions concern the cancellation and delay of holidays (Easter, Victory Day), the introduction of the digital permit system in Moscow and fines. In week 4, we observe a shift towards social support and business support, other week 3 topics remain. In week 5, the discussions touch on social protests and rallies (in Vladikavkaz and online), the extension of non-working days until the end of the May holidays and public restrictions in May. In week 6, people discuss the empty capital, the self-isolation regime extension, and governmental support. Week 7 concerns the situation in Moscow.

In week 8, the attention is drawn by the easing of the quarantine regime, work commencement, governmental measures, lifting restrictions, and antibody testing. In week 9, there is a mixture of topics: self-isolation, work, staying home, testing, Moscow region, digital permits, testing for antibodies. Week 10 discusses a possible lifting of restrictions in June and the upcoming celebration of the postponed Victory Day.

In general, the topic remains quite stable throughout the entire period, and clearly reflects the reaction to external changes and introduced measures. In the middle of the period, the discourse on the imposed restrictive measures in Moscow becomes more prominent, which may be due to the fact that it was in Moscow that some of the most stringent measures were introduced and subsequently imposed at the regional level. In the discussions, the sequence Moscow - Moscow oblast - regions is observed.

**Work**. The topic appears locally in weeks 2-3 and then reappears in week 6. In the first period, it is related to the start of non-working days and in week 3 it is fed by a discussion on remote work and learning. In week 6, the topic appears in a broader context - as a discussion of the prospects for starting work during or after the May holidays and salaries.

In general, the analysis of the topic content transformation confirms and expands the conclusions made during the examination of the dynamics of user activity within topics. The tendency of transition from a general broad context of week 1 to a personal perspective in weeks 2-4 is most clearly traced, followed by a return to a wide context with a gradual disappearance of personal discourses (even those that are characteristic of the second part of the period).

Also, the analysis of sub-discourses within topics shows that social networks respond quickly and clearly to all the main news feeds, and the reaction normally lasts one week. It is important to note that the topics that call for personal involvement of users keep being present in discussions for a longer time, even if they are generated by a news feed. Thus, the duration of the presence of the topic can serve as a signal of its importance for society and a fairly strong emotional response to it. However, it is important to remember that such an observation is not an unambiguous indicator and cannot be perceived as a sufficient condition. It requires further meaningful analysis and correlation with the news context of the relevant period.

3. Topic dispersion groups in activity clusters

The observed user behavior in clusters of activity confirms our statement about the spontaneous nature of this social network. Throughout the entire period, the number of accounts that wrote on one topic significantly exceeds the number of accounts in other groups: their ratio is above 70% in each week amid the overall sharp decrease in the number of accounts and contributions. Moreover, a large part of these accounts is usually in the temporal activity cluster, which means that their activity slows down or stops completely after the corresponding week. Along the considered period, the percentage of this group in the temporal cluster drops from 53% (week 1) to 32% (week 10) due to the overall activity decline while in the cluster of users with uniform activity it increases from 23% to 49%. Moreover, the number of different dispersion groups drops from 10 to 4 meaning that the accounts from other groups move to the monothematic group, which confirms our conclusion on the collapse of topic diversity by the end of the considered period.

## 5. Discussion and Conclusions

The proposed tool for the analysis of SM data complements the traditional approach, since the achieved results not only agree with the conclusions made in survey-based studies, but also provide additional data. Together with a visualization of adaptation mechanisms it expands the understanding of the process of adaptation to new conditions, since the processed texts are the product of a collective search for psychological meanings in the new reality. It is important to note that the selection of documents for the given corpus was keyword-based. Therefore, the trend of a gradual decrease in the overall discussion intensity and number of topics can be explained by the shift of the audience focus either to offline communication or to other topics (probably tagged with other keywords) that are not covered by the corpus. However, this trend can also represent the gradual social adaptation to the lifestyle changes imposed by COVID-19: the only concern left after the adaptation period is the pallid statistics, which is similar to checking the thermometer behind the window while life is going on. Indeed, during the study period (10 weeks from March 22 to June 1) we observe the gradual narrowing of the focus from a set of multiple diverse topics related to the COVID-19 spread to one dominating topic - infection and death statistics. Along the study period the multi-topic space with several major topics alternating their dominance (discussions on the introduced national-level measures, introspection) and other significant topics (including economics, foreign affairs, healthcare, COVID-19 vaccine and tests) gets transformed into a monotopic one (at weeks 9 and 10) where over a half of the accounts publish posts related to only statistical reports. The transition process goes, in the first place, through the personal experience and adaptation to the introduced changes at a "micro level" (finding solutions to household and job-related problems). Then it turns to a wider range of topics related to domestic politics (business, economics, aid to healthcare professionals) and foreign affairs until it eventually collapses to one major discussion on statistics: over a half of users produce over 80% of weekly posts on statistics and new cases reports. In other words, we deal with a topic space transformation from multitopic to monotopic through personal adaptation followed by the perception and acceptance of a wider context (the new reality at the macro level). It indicates the displacement of stress caused by the imposed restrictions and high anxiety levels characterized by the absence of a clearly defined subject of concern [Zevnik, 2017], the overall chaos and fragmentation, which is reflected in the increased user activity and the prevalence of the topic "Reflections (introspection) and conversations".

Similar conclusions were made based on the psychological measurements of anxiety and depression levels made during the first two weeks after the introduction of quarantine measures in Argentina. After two weeks the anxiety levels decreased significantly while the depression levels persisted almost unchanged [Canet-Juric, 2020]. It may reflect the ongoing adaptation process though accompanied with a persisting negative emotional state. Further, overcoming anxiety proceeds (from the thematic perspective) by solving, in the first place, more understandable problems of establishing personal life, organizing work and learning in a distance format, which corresponds to the traditional way of overcoming anxiety - establishing a daily routine [Hiremath, 2020]. The next major change in the topic structure is the expansion of the topic space to cover country- and world-level context, which may indicate a gradual decrease in anxiety at the personal level and the search for support in awareness ("I am not alone") and understanding of the broader context of the problem, which constitutes the next stage in overcoming stress [Kar, 2021].

The measurements of user activity in the first weeks after the announcement of the virus control measures (in the first Address to the Nation on 25.03.2020) performed in our study show that the user activity dynamics coincides with the dynamics of stress levels published in psychological research [Rusch, 2021: 18]. So, in the second week after the introduction of the most significant restrictive measures, there is a sharp increase in the number of the corresponding posts and the overall activity followed by a gradual decline along the weeks 4-6. It may indicate the decrease in the interest rate and relevance of these problems for users. Thus, the proposed approach can be used as a complementary tool and, possibly, an alternative to traditional psychological and sociological methods.

Appendix A. Descriptions of the manually consolidated topics and the main keywords that helped to distinguish between these topics.

| Topic | Assignment logic | Keywords |
| --- | --- | --- |
| Domestic politics | everything related to the President, the Government and the Addresses to the Nation | Putin, president, Vladimir, Mishustin |
| Healthcare professionals (incl. payments) | discussions on the work of healthcare professionals during the COVID-19 and assistance to them (financial aid in the first place) | physician, hospital, doctor, aid, receive, payment |
| Foreign affairs | everything related to foreign countries. This theme gathers all the interactions with the outside world (outside Russia), both political and economic, as well as the news about the situation in foreign countries. These are mostly discussions of foreign news stories that do not concern Russian citizens, therefore, they are combined into one theme. | names of foreign countries |
| Information sources | discussions of the information sources: news channels, LJ and Twitter accounts, news retelling | internet, online, read, channel, news, information, journalist, newspaper, account (several of these keywords should co-occur in a given post) |
| COVID (investigation, tests, treatment) | biomedical research and tests of COVID-19 vaccines, symptoms and their comparison to flu and pneumonia | vaccine, antibody, test(ing), sars, virus, science, laboratory, analysis |
| Hospitalization | hospitalization and treatment of COVID-19 in Russian health facilities | physician, hospital, patient, treatment, medicine, artificial lung ventilation |
| Folk_medicine, mysticism, conspiracy | fighting the virus with folk remedies and association of COVID-19 origin with mysticism and conspiracy theories. This group gathers esoterically oriented people prone to believe in the above things who both read and | garlic, ginger, besogon (a program about conspiracy theories), gates, bill, khodos (Ukranian personality) |

| | produce the corresponding content | |
|---|---|---|
| Virus_control_measures (personal level) | individual prevention and control measures. This theme is separated from the general (public level) measures and paid special attention, because, according to the results of psychological studies on this subject, the use of individual protective means is associated with the decrease in anxiety levels | mask, glove, protective, means, sanitizer |
| Virus_control_measures (public level) | general virus control measures introduced at the state level. In the first place, they include the introduction of the self-isolation regime in certain regions (Moscow and Moscow oblast), its variations and compliance with it in other regions, the rules for leaving home, the cancellation of festive events, and gradual restriction lifting | quarantine, permit, pass, go out, home, self-isolation, confinement, restrict |
| Solutions_to_household_problems | A small theme that gathers posts on the first response to the introduction of prevention measures and describes the related household problems and solutions to them | shop, go out, home, deliver, food, goods, buckwheat, toilet paper (the last two words occurred in a significant number of humoristic posts on buckwheat and toilet paper shortage in supermarkets) |
| Holidays | festive days (the Victory, day, Easter, Qurban Bayram) | bayram, day, victory, easter |
| Work | due to the introduction of non-working days, remote working and other work management measures, we decided to give prominence to the work topic (at the personal level) | work, money, employee, work, receive payments, get paid, salary, business |
| Entertainment_and_leisure | everything related to the leisure options during the lockdown (online activity, sports news, and a subtopic about church attendance) | free, online game, channel, book, video, access (here, each leisure type has its specific keywords, therefore, the assignment is made based on the co-occurrence of several terms) |
| Reflections_conversations | non-specific subtopics containing random reflections and discussions. They can be | know, talk, understand, criticize, want, panic, fear, sense (here, keywords are difficult to |

|  | considered as an alternative to real communication that was limited during the lockdown | define in terms of topic focus, however, the most representative ones are related to states (emotional state in particular) and actions) |
| --- | --- | --- |
| Regional_problems | several topics clearly associated with Russian regions (political subdivisions) | Nizhniy Novgorod, Tatarstan, Ingushetiya, Bashkortostan, Rostov (topological names) |
| Statistics (infection, death) | new cases, infection spread and death dynamics | spread, infection, new, case, number, infected, names of months |
| Economic_issues | topics concerned with both domestic and international economies. What these topics have in common is the "macro-level" (global) type of discussion that concerns country and world economic problems and does not touch on one's personal situation. This theme also includes the topic about small and medium business support in Russia. | economics, crisis, market, business, oil, price, finance, |

Appendix B. Twitter: the per-topic percentage of users and posts in week 1 (1) and week 2 (2), week 3 (3) where *accratio* stands for the percentage of users in a topic in a given week and *textratio* is the percentage of texts in a topic in a given week.

1) Week 1

2) Week 2

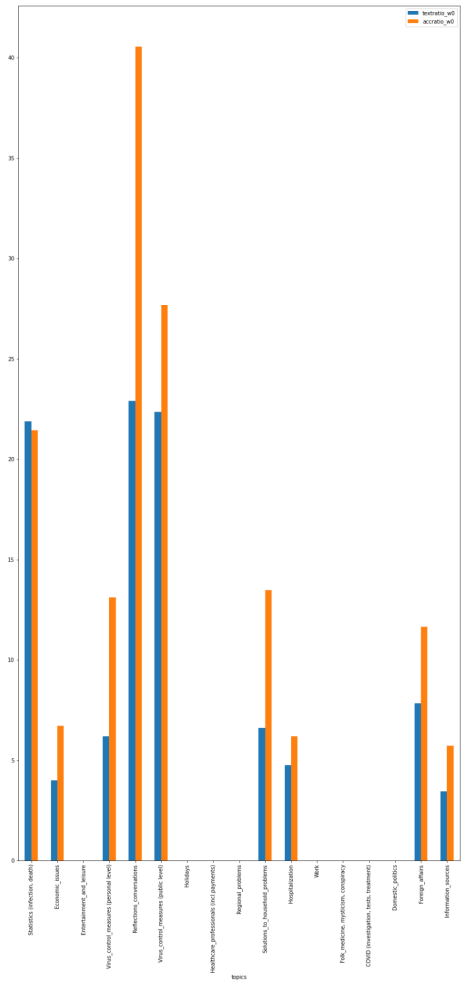

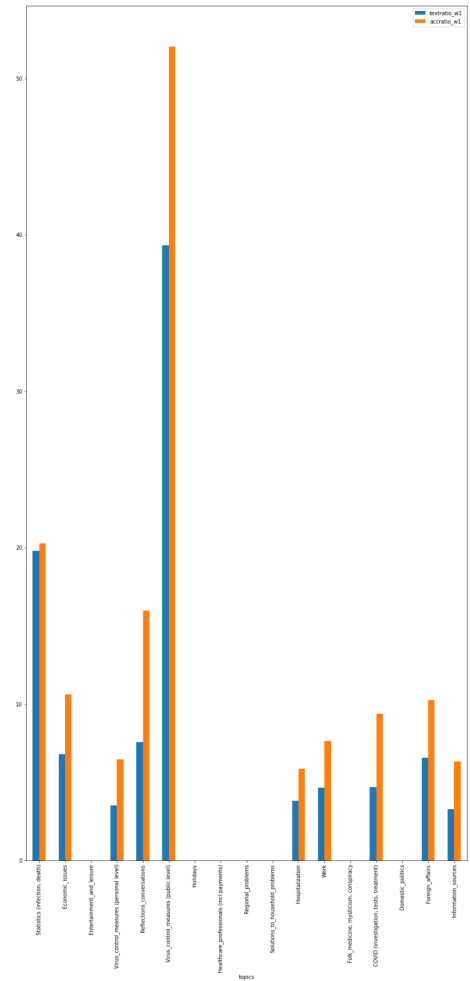

3) Week 3

4) Week 4

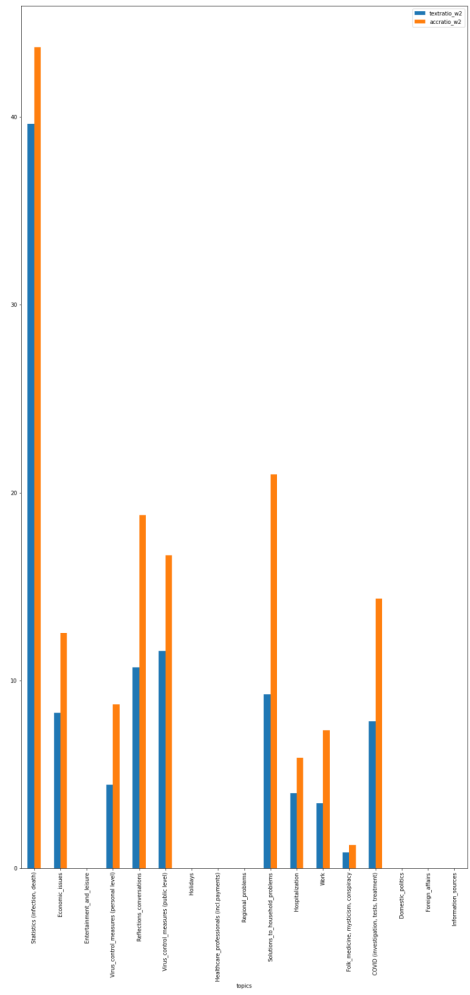

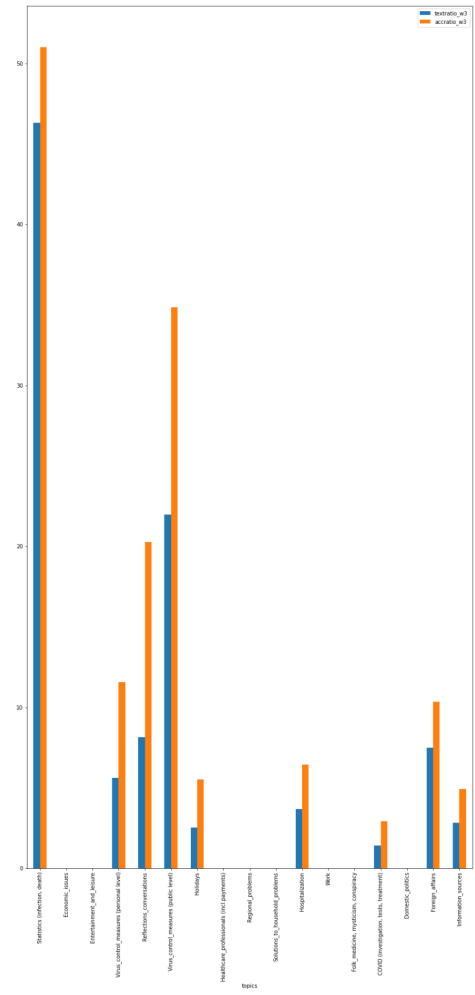

5) Week 5

6) Week 6

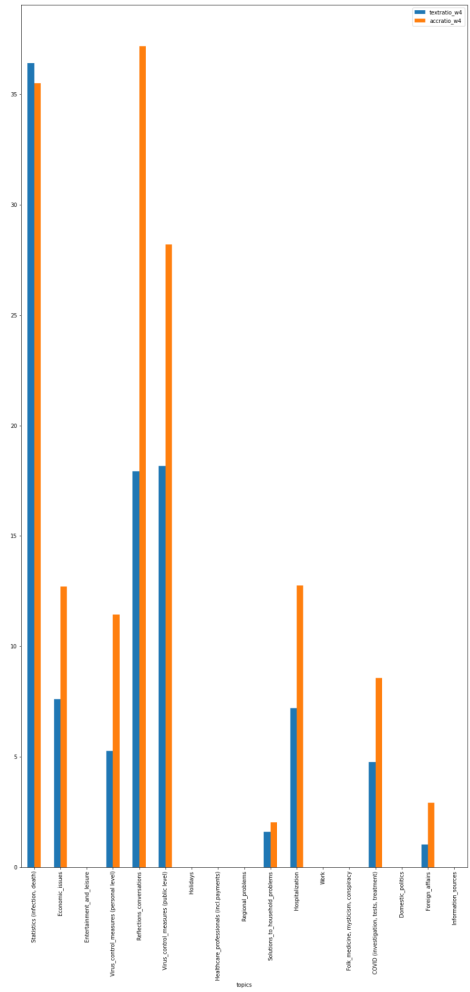

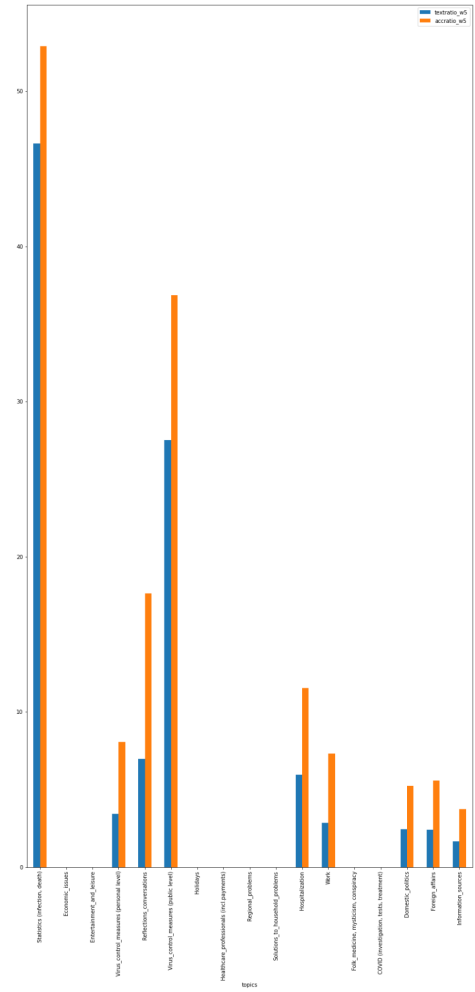

7) Week 7

8) Week 8

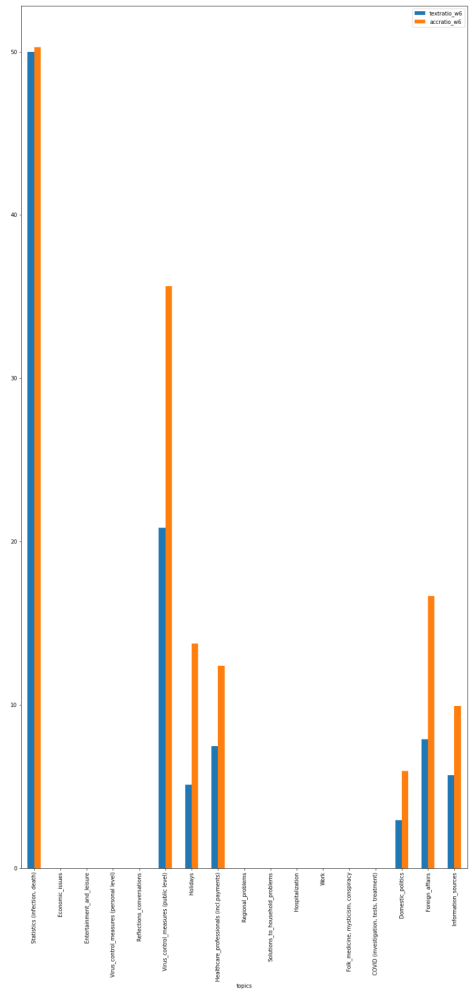

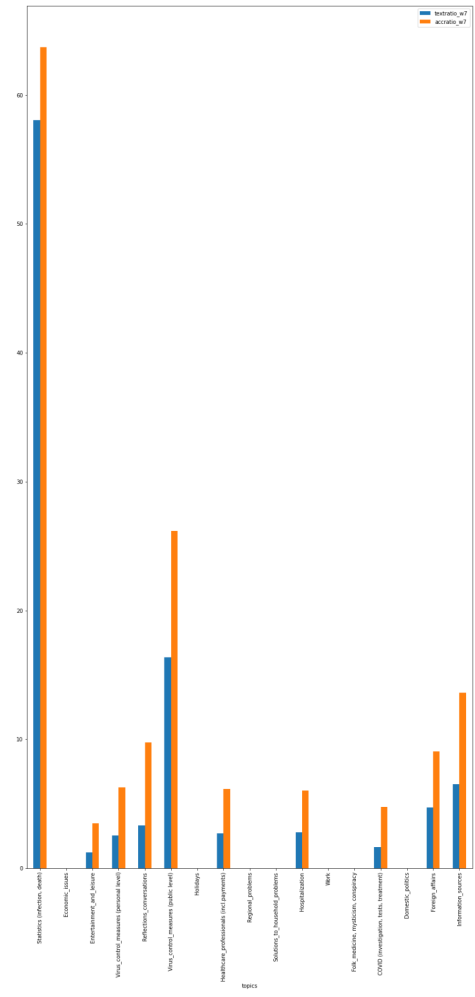

9) Week 9                    10) Week 10

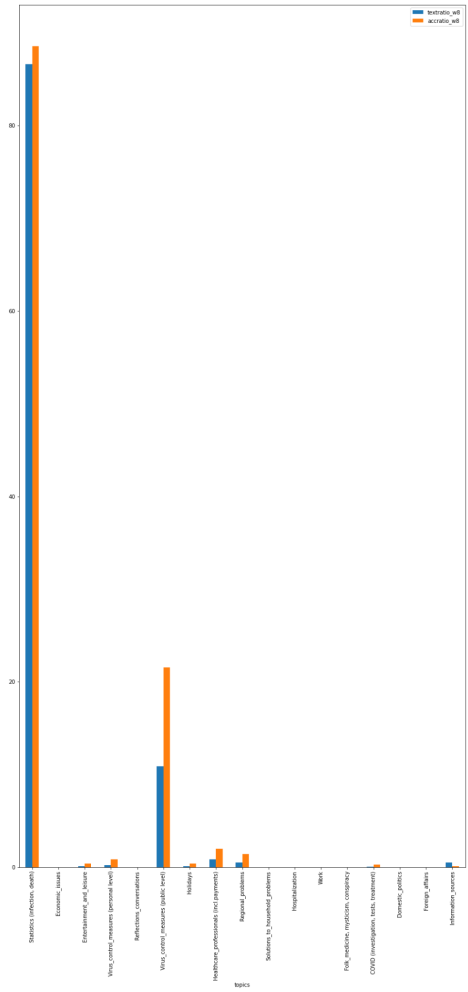
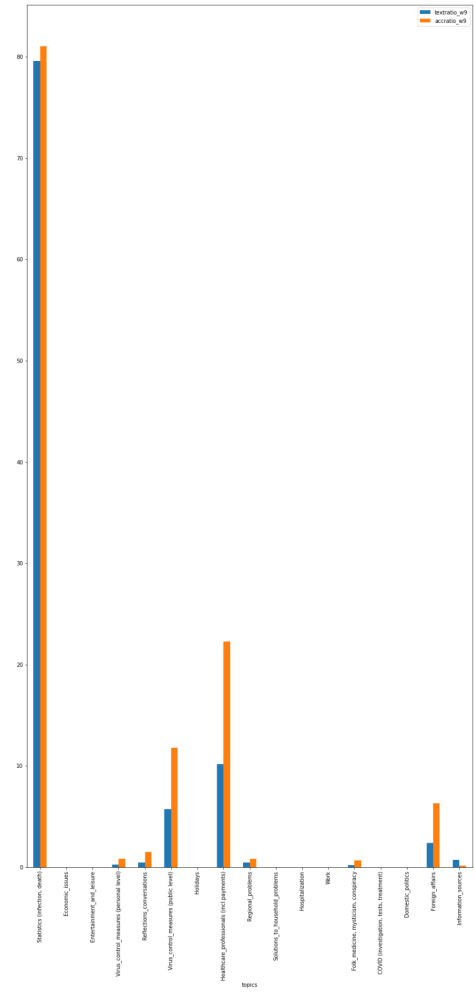